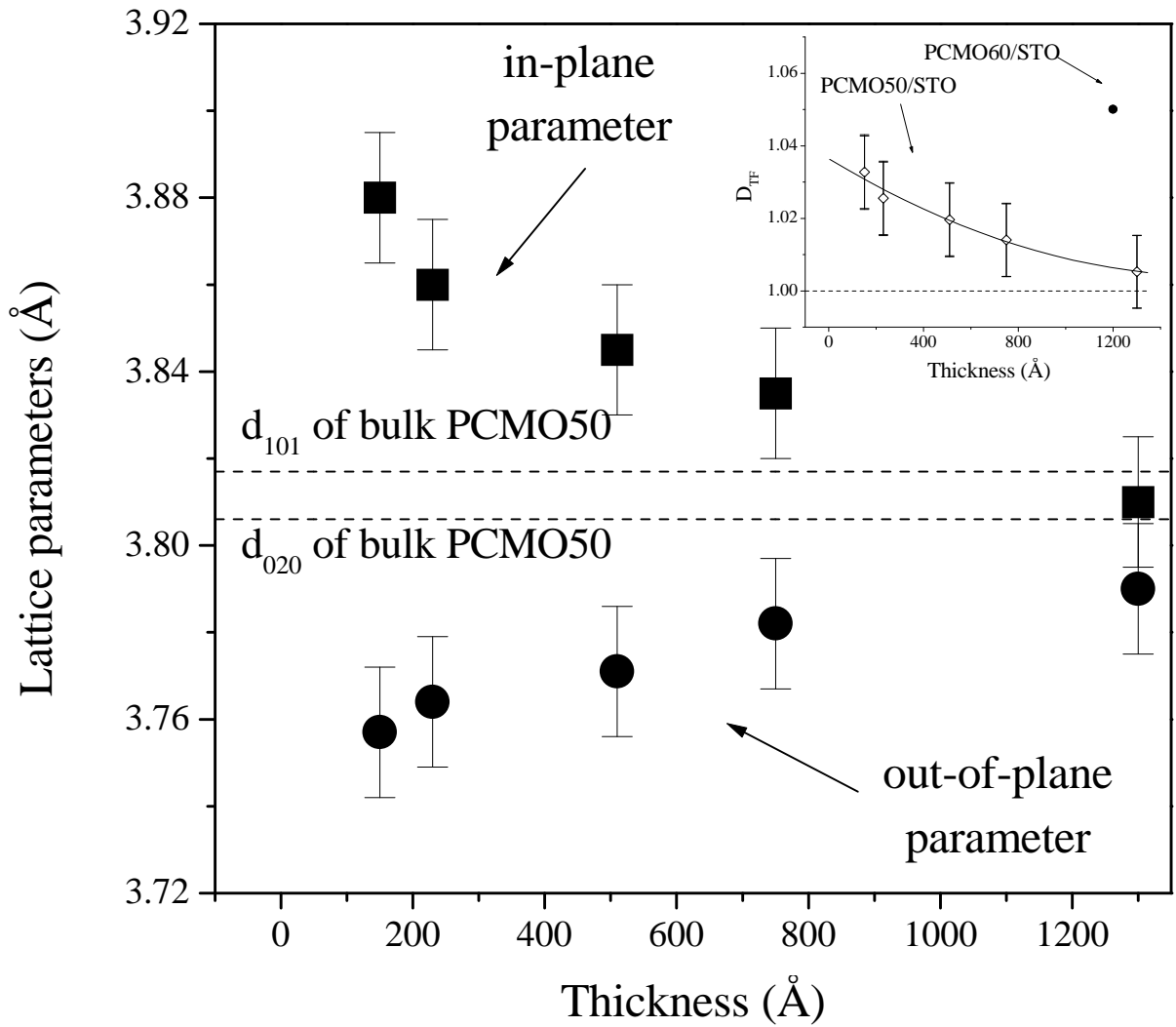

Fig.1

W. Prellier et al.

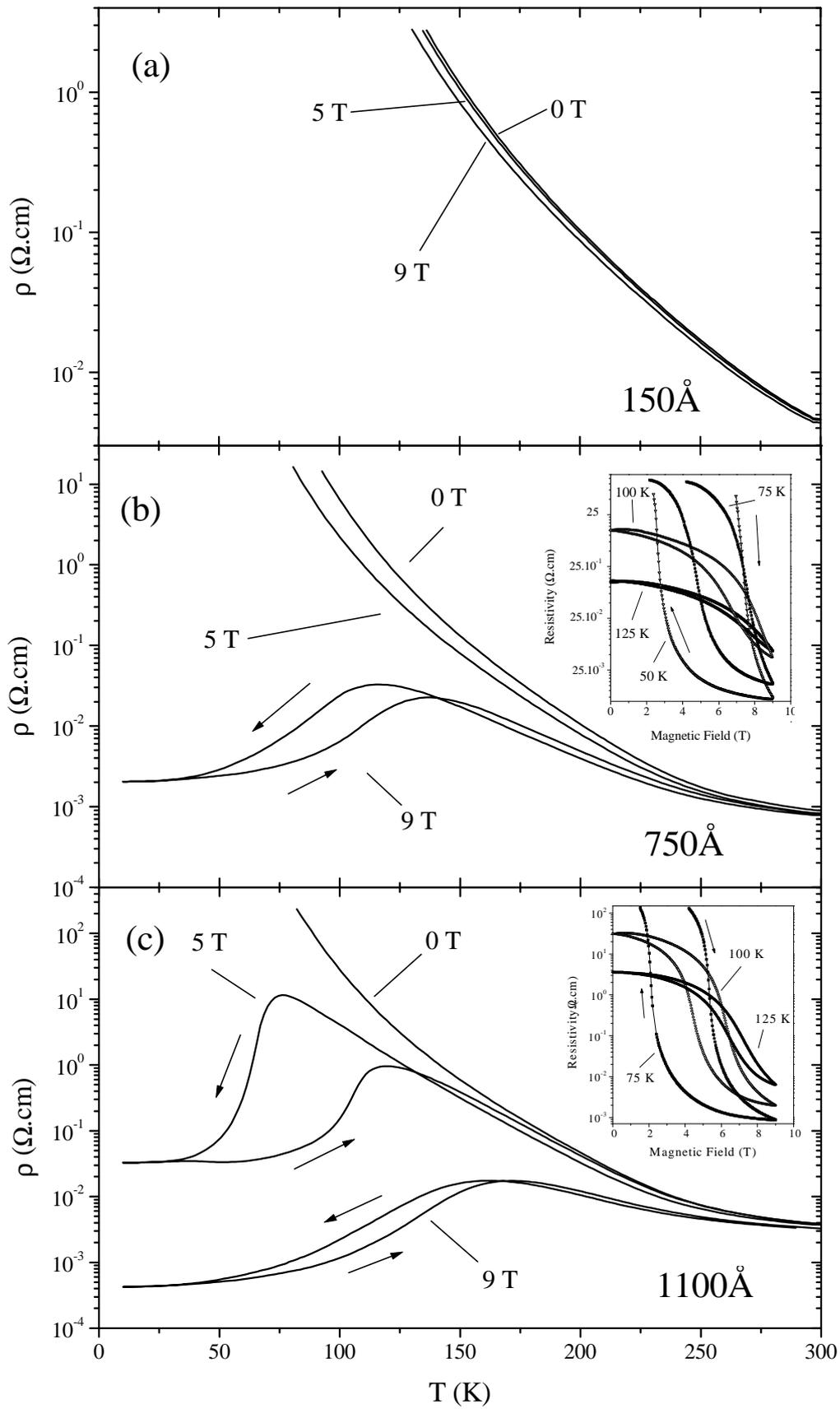

W. Prellier et al.

Fig.2

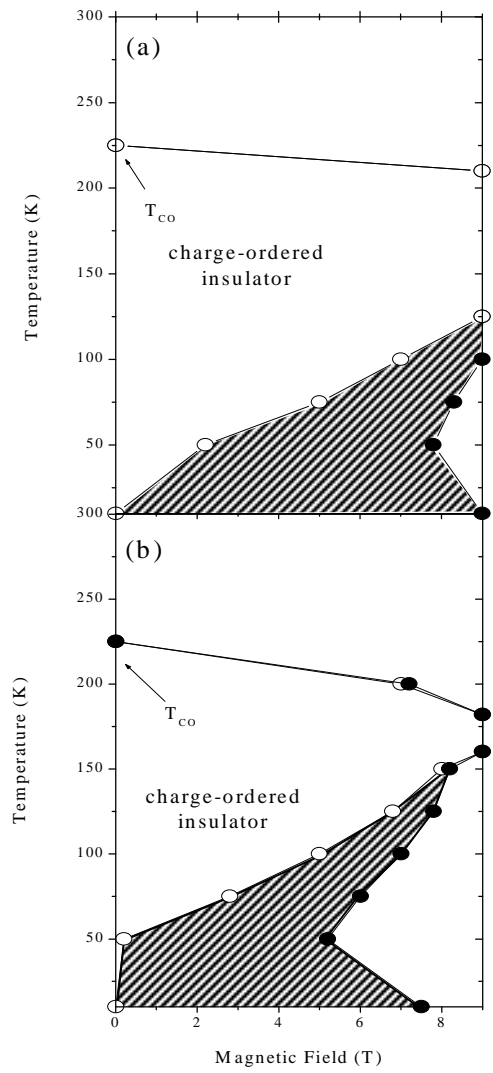

Fig.3

W. Prellier et al.

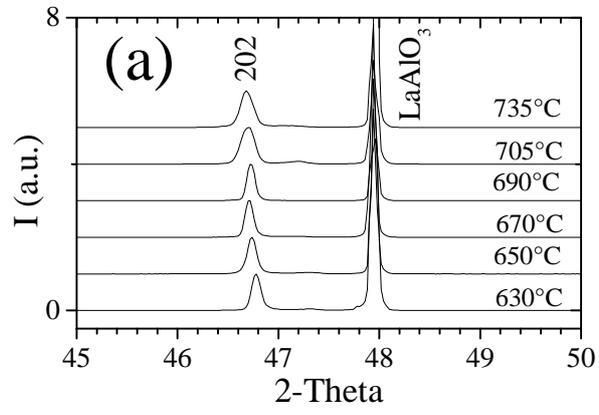

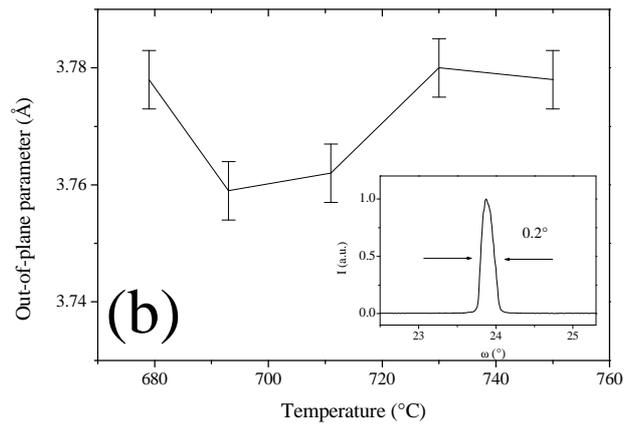

Fig.4

W. Prellier et al.

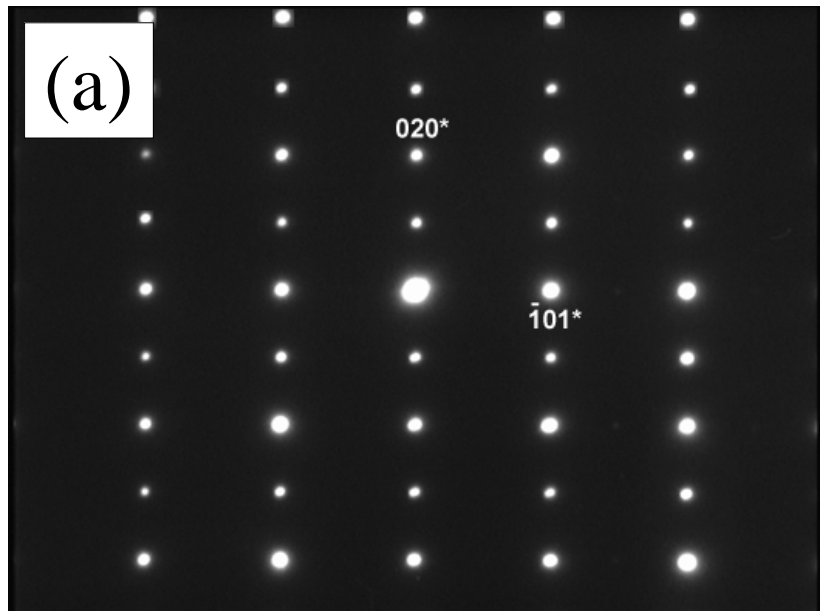

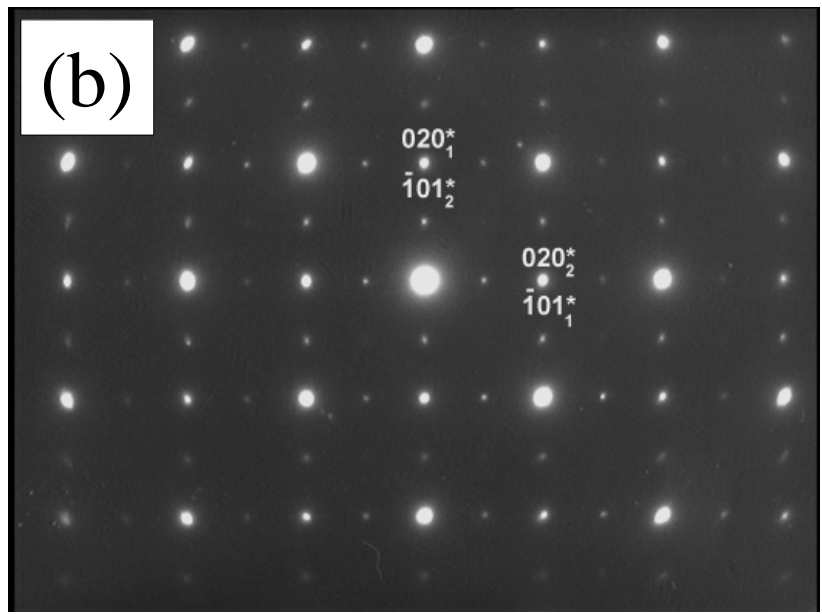

Fig.5    W. Prellier et al.

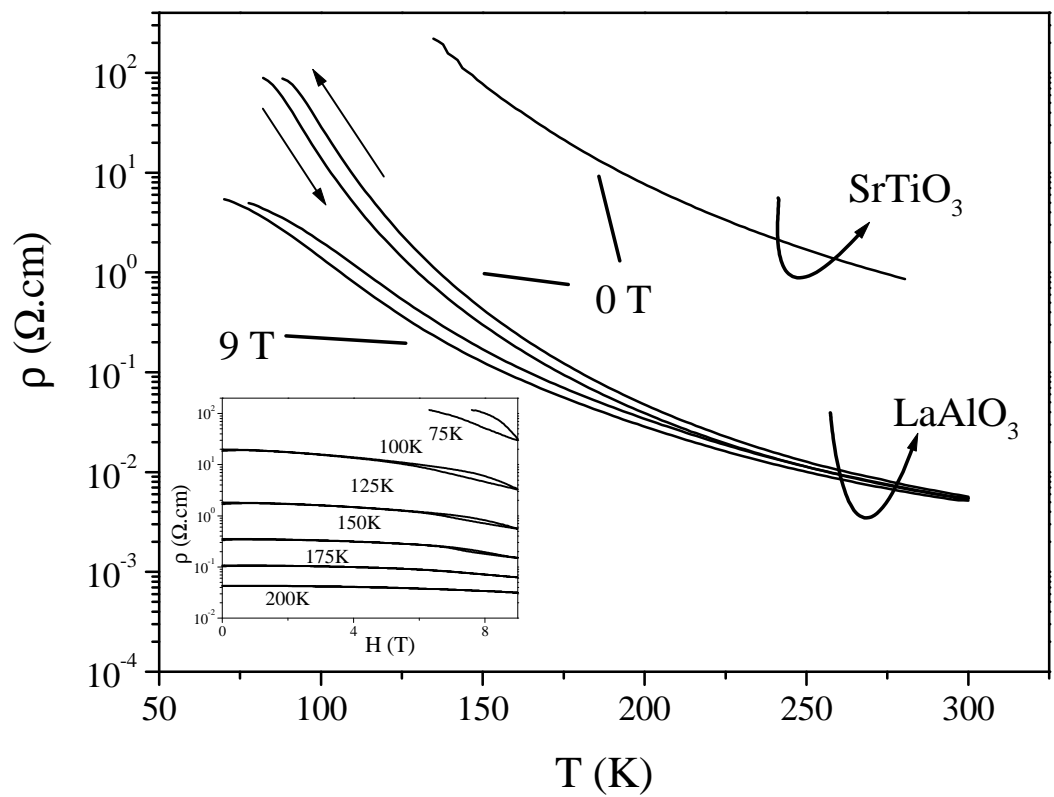

Fig.6

W. Prellier et al.

# Stress-induced metallic behavior under magnetic field in $Pr_{1-x}Ca_xMnO_3$ (x = 0.5 and 0.4) thin films

W. Prellier[1], Ch. Simon, B. Mercey, M. Hervieu, A.M. Haghiri-Gosnet, D. Saurel, Ph. Lecoeur and B. Raveau

Laboratoire CRISMAT, CNRS UMR 6508, Bd du Maréchal Juin, 14050 Caen Cedex, FRANCE

**Abstract**

We have investigated the role of the stress induced by the presence of the substrate in thin films of colossal magnetoresistive manganites on structural, resistive and magnetic properties. Because of the strong coupling between the small structural distortions related to the charge-ordering (CO) and the resistive properties, the presence of the substrate prevents the full developpement of the charge ordering in $Pr_{0.5}Ca_{0.5}MnO_3$, especially in the very thin films. For thicker films, the CO state exists, but is not fully developped. Correlatively, the magnetic field which is necessary to suppress the CO is decreased drastically from 25 Tesla to about 5 Tesla on $SrTiO_3$ substrates. We have also investigated the influence of the doping level by studying the case of $Pr_{0.6}Ca_{0.4}MnO_3$.

---

[1] Prellier@ismra.fr

The magneto-transport properties of the manganites perovskites $RE_{1-x}A_xMnO_3$ (RE and A are rare-earth and alkaline-earth ions, respectively) have attracted renewed interest because of the observation of colossal magnetoresistance (CMR), i.e. a huge decrease in resistance when applying a magnetic field [1-4]. Since most of the technological applications require thin films, it is essential to understand the effects of the substrate-induced strains on the properties of these manganites. These materials are very sensitive to the strains and even very small perturbations may result in observable effects on the properties. These effects have indeed been studied experimentally on various compounds [5-10]. Moreover, the study of CMR and related properties of $RE_{1-x}A_xMnO_3$ compounds has brought novel properties related to charge and dynamics in these oxides. Charge-ordering (CO) is one of it. This phenomenon is not new since it is known in $Fe_3O_4$ [11]. In manganites, the phenomenon of CO, first investigated by Jirak et al. by neutron diffraction [12] in $RE_{0.5}A_{0.5}MnO_3$, is probably one of the most remarkable effects. It appears for certain value of x and particular average A-site cation radius. It corresponds to a 1-1 ordering of the charges in two different Mn sublattices. In fact, the metallic state becomes unstable, below a certain temperature ($T_{CO}$), and the material goes to an insulating state. $T_{CO}$ decreases with increasing field and the insulating CO state can be totally suppressed (melted) by the application of an external magnetic field [13-15]. The CMR properties result from a competition between an antiferromagnetic (AFM) and ferromagnetic (FM) state. By the way, any factor able to destabilize the CO-AFM state is of interest.

In the bulk, $RE_{1-x}A_xMnO_3$ materials, several factors have been determined as important: average A-size cation, size mismatch, valence effect by varying the x value of the oxygen content, charge mismatch, nature of the Mn site doping element, magnetic field during cooling, electron irradiation…[16]. Working on thin films, another factor crucial is added, namely the strain effect [17]. However, taking into account all the key factors to understand the way the properties are modified is highly difficult because numerous parameters have to be controlled. It is the reason why we start to study $Pr_{0.5}Ca_{0.5}MnO_3$ (PCMO50). It was indeed shown that, in bulk, the CO is very sensitive to chemical and geometrical parameters [18]. A 25T magnetic field is required to melt the CO state in the bulk $Pr_{0.5}Ca_{0.5}MnO_3$ [19].

We have shown recently that, when this compound is synthesized as a tensile thin film on $SrTiO_3$ (STO), the melting magnetic field is reduced to 6T [20], whereas a value of 9T is not sufficient when the film is grown on $LaAlO_3$ (LAO) to collapse this state [21]. This means that uniaxial strain (1-2% in the plane) leads to dramatic changes as compared to the bulk material. One would expect that the relaxation of the strains along the direction normal to the plane of the substrate should also induce changes in the electronic properties.

Considering the lattice parameters, transport measurements and magnetization, we have studied the effects of the thickness upon the structural and physical properties of $Pr_{0.5}Ca_{0.5}MnO_3$ thin films grown on STO, using pulsed laser deposition. On the basis of these results, we have determined a temperature-field phase diagram for CO thin films and compared it to the bulk material. The effect of the substrate-induced strains and film thickness on the properties has been tested. A comparaison with the properties of $Pr_{0.6}Ca_{0.4}MnO_3$ (PCMO60) thin films on LAO and STO, is also reported in this communication.

Thin films of $Pr_{1-x}Ca_xMnO_3$ were grown in-situ using the Pulsed Laser Deposition (PLD) technique. The targets used had a nominal composition of $Pr_{1-x}Ca_xMnO_3$ (x=0.5 and 0.4). The substrates used were (100)-$LaAlO_3$ (LAO), which has a pseudocubic crystallographic structure with a=3.788Å, and (100)-$SrTiO_3$ (STO), cubic with a=3.905Å. The laser energy density on the target was about 1.5J/cm2 and the deposition rate was 4Hz. The substrates were kept at a constant temperature around 700°C during the deposition, which was carried out at a pressure of 300mTorr of oxygen. After deposition, the samples were slowly cooled to room temperature at a pressure of 400Torr of oxygen. Further details of the target preparation and the deposition procedure are given elsewhere [20-21]. The structural study was carried out by X-Ray diffraction (XRD) using a Seifert XRD 3000P for the $\Theta$-$2\Theta$ scans and a Philips MRD X'pert for the in-plane measurements (Cu K$\alpha$, $\lambda$=1.5406Å). The in-plane parameters were obtained from the $(103)_P$ reflection (where p refers to the ideal cubic perovskite cell). A JEOL 2010 electron microscope was used for the electron diffraction (ED) study. Resistivity ($\rho$) was measured by a four-probe method with a Quantum Design PPMS (up to 9T) and magnetization (M) was recorded using a Quantum Design MPMS SQUID magnetometer (up to 5T) as a function of the temperature (T) and the magnetic field (H)

*$Pr_{0.5}Ca_{0.5}MnO_3$ thin films : thickness dependence*

Fig.1 shows the evolution of the in-plane and out-of-plane lattice parameters vs. thickness for PCMO50 deposited on STO. As already shown, films grown on STO are [010]-oriented, refering to the parameters of the *Pnma* structure [20] {Wil2}. It means that the [101]* and [10$\bar{1}$]* directions of the film (in-plane parameters) are parallel to the a* and b* of STO whereas the [010]* of the film is perpendicular to the substrate (out-of-plane parameter). For the very thin film (150Å), the in-plane parameter is nearly equal to the one of the substrate (3.88Å vs. 3.9Å for STO). When the thickness of the film increases, the in-plane parameter decreases whereas the out-of-plane parameter concomitantly increases. For large thickness (~1300Å), both lattice parameters tend towards the bulk values. However at 1400Å, they do not reach exactly the bulk values and the film is not fully relaxed. Note that the cell volume is preserved constant and almost equal to that of the bulk (~56Å$^3$ referring to the perovskite cell). Considering the parameters of the bulk compound (a=$a_P\sqrt{2}$ =5.395Å, b=$2a_P$=7.612Å and c=$a_P\sqrt{2}$ =5.403Å [12]), the distances $d_{101}$ and $d_{020}$ are respectively calculated to be equal to 3.817Å and 3.806Å. Thus, regarding the lattice parameter of STO and the film orientation, the film is under compression. Moreover, the temperature dependence of these cell parameters should be very different from that of bulk samples since they are constrained by the substrate. It was shown previously that the orthorhombic distortion induced by the CO ordering is, at low T, strongly reduced by the presence of the substrate, preventing the perfect lock-in of the q value of the CO to the ideal value 0.5 (q=0.45 is for the thicker film [20]). Fig.2 shows the ρ(T) curves registered under different applied magnetic fields (0, 5T and 9T) for three film thicknesses. In the absence of applied field, a semiconducting behavior is observed whatever the thickness with an anomaly around 225K, corresponding to the $T_{CO}$ (this value is close to that found in the bulk material [19]). On the contrary, the results are drastically different under application of external magnetic fields. A field up to 9T (our maximum value) has almost no effect on the thinner film (150Å): the film is still semiconducting and the magnetoresistance is very small (Fig.2a). For an intermediate thickness (750Å), while a 5T magnetic field keeps the film semiconducting, a 9T field renders it metallic with a insulator-to-metal transition ($T_{MI}$) close to 125K (see Fig.2b). This feature is a characteristic of the melting of the CO state. On the thicker film (1100Å), the melting magnetic field is reduced to 5T (Fig.2c). The thickness dependence of the properties suggests that the strains play an important role in determining the melting magnetic field. When the thickness of the film is increasing, there is a reduction of the strain on the film. However, these results show that the properties of the film do not approach those of the bulk as the substrate-induced strain reduces. In single crystal $Pr_{0.5}Ca_{0.5}MnO_3$, with a 6T magnetic field, the material remains insulating [22], whereas the 1100Å film becomes metallic. Note that all the measurents done at this stage are averaged. This means that for example, the lattice parameters are an averaged value in the overal film. It is possible that the first layers, close to the substrate-interface, are more strained that the upper ones. Thus, such variations could induce small difference in the melting magnetic field. But the viewing our results on $Pr_{0.5}Ca_{0.5}MnO_3$ grown on $LaAlO_3$ [21] where the film did not show extended defects or large lattice variations, we expect to have the same scenario in the films deposited on $SrTiO_3$. A detailed transmission electron microscopy study will be necessary to ensure this statement.

The magnetic field dependence of the resistivity at various temperatures is respectively shown in the inset of Fig.2b and Fig.2c for the 750Å and the 1100Å thick films. We are not able to measure the ρ(H) dependence for the 150Å film due to the limitation of our apparatus (9T). In both cases, the resistivity shows an important decrease on a logarithmic scale at a critical field ($H_C$) indicating the field-induced melting of the CO state. This field-induced insulator-to-metal transition is correlated to the CO melting and takes place below $T_{CO}$. The critical fields in the field increasing scan and in the field decreasing scan are respectively represented by $H_c^+$ and $H_c^-$. This decrease of the resistivity vs. magnetic field can be viewed as CMR of about five orders of magnitude at 75K (inset of Fig.2c). A clear hysteresis is seen at these temperatures as previously reported on several CO compounds [14,22]. This hysteretic region is more pronounced when the temperature is decreasing. The temperature dependence of the large hysteresis region is a feature of a first order transition and has been extensively studied for the composition $Nd_{0.5}Sr_{0.5}MnO_3$ in ref. [14]. From the resistivity measurements, we can deduce the phase diagrams for the 1100Å film (Fig.3a) and for the 750Å film (Fig.3b). Two points must be emphasized. First, as already mentioned, the transition of the CO state (i.e. a field-induced metallic state) is easier or requires a lower field in case of a film than in the bulk [22] (5T for a 1100Å film and 9T for the 750Å one). Second, the shape of the H-T phase diagram is totally different as compared to the bulk single crystal PCMO50. $H_C$ is smaller than 9T at low temperature, then $H_C$ is decreasing when T is increasing. Above 50K, $H_C$ begins to increase again to 8T. In the mean time, the hysteretic region is shrinking. Note that this region is larger in the case of the 750Å film indicating that the CO state is more stable as previously seen from the ρ(T) curves (Fig.2). The important new result is the thickness dependence of the film properties. The distortion induced by the CO cannot fully develop due to the strains imposed by the substrate. Consequently, the CO is limited to an incommensurate value leading to a less stable CO state, more sensitive to the field.

For describing and assessing the structural distortion of the perovskites, different parameters are commonly accounted, as the tolerance factor of Goldsmidt, the notation of Glazer or the D factor of Poppelmeir. Since the substrate induces a specific strain to the thin film, one considered another $D_{TF}$ parameter to follow this lattice distortion, defined as the ratio between the in-plane and the out-of-plane parameters. The first factors (t and D) vary with the cation and anion contents. On the opposite, the variation of $D_{TF}$ will be considered, for a given composition, as a fucntion of specific parameters, i.e. substrate parameters and film thickness. Moreover, the $D_{TF}$ expression varies with the film orientationn, namely $D_{TF} = d_{101}/d_{020}$ for a [010]-oriented *Pnma*-type film (case of PCMO50/STO) and $D_{TF} = d_{020}/d_{101}$ for a [101]-oriented one (case of PMCO50/LAO). The evolution of $D_{TF}$ vs. thickness is plotted in the inset of Fig.1 for PMCO50/STO. From this graph and Fig.2, it clearly appears that the critical field required to melt the CO state is decreasing as the distorsion of the thin film decrases.

*Pr$_{0.6}$Ca$_{0.4}$MnO$_3$ thin films*

To better understand the effects of strains, we looked, in a second step, for a composition susceptible to exhibit a melting magnetic field significantly low (down to 6T) when the film would be grown under tensile stress. Considering the Pr$_{1-x}$Ca$_x$MnO$_3$ diagram, we decided to grown a film of Pr$_{0.6}$Ca$_{0.4}$MnO$_3$ [22]. We synthesized this material on both LAO and STO, keeping a constant thickness chosed to 1200Å (i.e. the thickness where PCMO50 on STO exhibits the lower melting magnetic field up to now). Fig.4a shows the XRD in the range 45-50° at room temperature for PCMO60 films grown on LAO at different temperatures. One peak appears close to the substrate indicating a perovskite-like structure with an out-of-plane larger than the one of LAO. One notes that the lattice parameter shifts towards the high value as the temperature is increasing. However, for 690°C the peak is symmetrical and sharp indicating a high quality of the film, which is confirmed by the value of the full width at half maximum (FWHM) of 0.3° (not shown). On STO, the scenario is almost the same as the films grown on LAO; the lattice parameter is changing vs. the deposition temperature (Fig.4b). However, in this case, the parameter presents a minimum for a deposition temperature around 690°C leading to an out-of-plane parameter close to 3.76Å. At this temperature, the film is well crystallized and highly oriented (see the value of the FWHM of the rocking-curve less than 0.2° in the inset of Fig.4b). Considering the bulk parameters (*Pnma*, with is a=5.415Å, b=7.664Å and c=5.438Å ) [12], we calculated that the distances d$_{101}$ and d$_{020}$ are very close (3.83Å), making difficult to differentiate one from the other. Two comments should be done. First, in these films from our XRD measurements, we cannot find out the orientation regarding the substrate plane. The ED study will give us the final answer. Second, the cubic LAO substrate has a smaller lattice parameter than PCMO60 (3.79Å vs. 3.76Å) whereas STO has a bigger lattice constant (3.9Å vs. 3.76Å) at room temperature. Thus, films on LAO are compressed whereas on STO films are expanded (in the direction normal to the substrate plane).

Several areas of the film were characterised by ED to investigate the orientation of the film. The cell parameters of the film are a = a$_p\sqrt{2}$ , b = 2a$_p$, c = a$_p\sqrt{2}$ , similarly to that of the bulk structure. A typical ED pattern of a PCMO on LAO film is given in Fig.5a. The film is [101]-oriented as previously reported for several manganites deposited on LAO (with a GdFeO$_3$-type structure [21,23]). On STO, the orientation is different since the film is [010]-oriented (Fig.5b). Resistivity measurements of PCMO60 grown on LAO and STO with and without magnetic field are shown in Fig.6. The magnetic field is applied in the plane of the substrate plane. On both substrates, the films show an insulating behavior without a magnetic field. We do not clearly observed a small shoulder (usually associated to $T_{CO}$) in the ρ(T) at a temperature around 240K like in the bulk material [22]. The application of a 5T magnetic field has no effect on the film behaviours. However, at low temperature, the PCMO60 on LAO exhibits a small decrease in resistivity, applying an external magnetic field of 9T (see the resistivity field dependence in the inset of Fig.6). In spite of this, the film is never totally metallic and no metal-insulator transition is observed in this substrate. On STO, the film remains insulator with no CMR effect, whatever the magnetic field up to 9T (limits of the apparatus). Magnetic measurements performed on both samples reveal no transition since we were not able to go higher than 5T (limits of our measurements). In-plane measurements were carried out on the 1200Å thick films and we respectively found a parameter of 3.7402Å on LAO and 3.9502Å on STO. In PCMO60, the distortion $D_{TF}$ (as defined previously in the paper) is thus calculated to be respectively (~0.99 on LAO and ~1.05 on STO (see the inset of Fig.1).

These values are drastically different from the ones obtained in the bulk compound (~1.001) (calculated by the ratio between the $d_{020}$ and the $d_{101}$ [12]). This implicates that the enhancement of the lattice distortion suppresses the charge ordering in thin films. Since the film is less distorted on LAO (the distortion is closer to 1 than in the case of STO), we are able to see a small decrease of the resistivity with a magnetic field of 9T. On STO the film is much more distorted than the bulk preventing the collapsing of the charge-ordering state with a magnetic field around 7T. In fact, these results are in agreement with the studies on PCMO50, where the less distorted film has a lower melting magnetic field.

In summary, we have firstly shown that the melting magnetic field of the insulating CO state, in tensile PCMO50 films, is strongly sensitive to the thickness of the film. While, the application of a magnetic field up to 9T has almost no effect on the resistivity of a very thin film, a 5T magnetic field can completely melt the CO state of a 1100Å thick film. Secondly, we have synthesized PCMO60 films on LAO and STO. Our results reveal that the film on LaAlO$_3$ has a lower melting magnetic field. In both cases, the reduction of the critical magnetic field can be explained by the distortion of the cell of the film. Finally, it appears that a decrease of the cell distortion, due to either an increase of the thickness in thin films under tensile strain, or a strained film, plays a role similar to a decrease of the A-site cation size in bulk material. For example, the stability of the AFM-CO state is decreasing in $Ln_{0.5}Ca_{0.5}MnO_3$ when Ln goes from Y to La [24]. Moreover, this work highlights also an important point of the magnetotransport of manganites thin films regarding the lattice distortion. As shown, drastic distortions are induced by the substrate, from pseudo-tetragonal for PCMO50 [010]-oriented on STO to monoclinic for PCMO50 [101]- oriented on LAO. Despite these strong extension and compression, the structural type remains GdFeO3-type, implying that the tilt and the rotations of the octahedra are similar. It undoubtedly imposes strong variations of the Mn-O distrances and O-Mn-O angles, which are so important for the CMR properties.

**Figure Captions**

Fig.1: Evolution of the in-plane [101] (i.e. $100_P$), and the out-of-plane parameter [020] (i.e. $001_P$) of PCMO5/STO films with different thickness at room temperature. The inset depicts the evolution the distorsion $D_{TF}$ for PCMO50/STO films vs. thickness. Full dot corresponds to the distorsion calculated for the 1200Å thick films of PCMO60 on STO. The dashed lines indicate the values of the bulk lattice parameters PCMO50.

Fig.2: $\rho(T)$ under different magnetic field (0, 5T and 9T) applied in the plane of the substrate for different PCMO50 films on STO (a): 150Å, (b): 750Å and (c): 1100Å. Arrows indicate the direction of the temperature variation. Experimental points were taken with a stabilized temperature and a control of the sample film itself. The crossing of the curves ramping up and down in temperature is real. The inset of Fig.2a and Fig.2b respectively depict the $\rho(H)$ for a 750Å and 1100Å film. Runs in field-increasing and field-decreasing are denoted by arrows.

Fig.3: Electronic phase diagram of PCMO50/STO film determined by the critical fields. $H_c^+$ (open dots) and $H_c^-$ (filled dots) are respectively taken as the inflexion points in $\rho(H)$ for up and down sweeps. In the dashed region, both CO insulating and metallic state are coexisting. (a): 1100Å (b): 750Å.

Fig.4: (a) Room temperature XRD of PCMO60 on LAO. (b):Evolution of the out-of-plane parameter of a PCMO60 film grown on STO. The inset shows a rocking-curve recored around the main peak.

Fig.5a: ED pattern of a PCMO60/LAO LAO taken along the [101] direction .

Fig.5b: ED pattern of a PCMO60/STO taken along the [010] direction.

Fig.6: $\rho(T)$ as a function of the field for a PCMO60 1200Å thick film on LAO and STO. The inset depicts the $\rho(H)$ for the film grown on LAO.